\documentstyle[12pt]{article}
\begin{document}

\title{\Large\bf Mass generation for gauge fields in the
Salam-Weinberg theory without Higgs}

\author{J. Barcelos-Neto\thanks{\noindent e-mail:
barcelos@if.ufrj.br}~~ and S. Rabello\thanks{\noindent e-mail:
rabello@if.ufrj.br - Address after February 19th, 1996: Department of
Physics, Stanford University, Stanford, CA94305-4060, USA}\\ 
Instituto de F\'{\i}sica\\ 
Universidade Federal do Rio de Janeiro\\ 
RJ 21945-970 - Caixa Postal 68528 - Brasil\\
\\
\\}
\date{}

\maketitle
\abstract
We consider the Salam-Weinberg theory by introducing tensor gauge
fields. When these fields are coupled in a topological way with the
vector ones, the resulting system constitutes an alternative
mechanism of mass generation for vector fields without the presence
of Higgs bosons. We show that these masses are in agreement with the
ones obtained by means of the spontaneous symmetry breaking.

\vfill

\vspace{1cm}
\newpage

{\bf 1.} The origin of mass generation for gauge fields in the
Salam-Weinberg~(SW) theory has been an interesting and intriguing
problem since its proposal. Nowadays, it is widely accepted that
spontaneous symmetry breaking together with the Higgs mechanism is
the most probable explanation for the origin of these masses.
However, if this is actually true, the Higgs bosons must exist in
nature. The point is that there is no precise theoretical prediction
on the mass scale where these fields could be found and experiments
till now have shown no evidence about them.

\medskip
More recently, it has been pointed out that a vector-tensor gauge
theory~\cite{Allen}, where vector and tensor fields are coupled in a
topological way by a kind of a Chern-Simons term, constitutes an
interesting mechanism of mass generation for vector fields that is
not plagued with Higgs. The general idea of this mechanism resides in
the following: Tensor gauge fields~\cite{Kalb} are antisymmetric
quantities and consequently in $D=4$ they exhibit six degrees of
freedom. By virtue of the massless condition, the number of degrees
of freedom goes down to four. Since the gauge parameter is a vector
quantity, this number would be zero if all of its components were
independent. This is nonetheless the case because the system is
reducible and we mention that the final number of physical degrees of
freedom is one. When the Chern-Simons term is introduced, where
vector and tensor field are coupled in a topological way, this
remaining tensor degree of freedom can be absorbed by the vector one
to acquire mass~\cite{Allen,Barc1}. We mention that this peculiar
structure of constraints implies that quantization deserves some care
and a reasonable amount of work has been done on this
subject~\cite{Kaul}.

\medskip
The purpose of our paper is to use this mechanism in the SW theory in
order to generate mass for the weak gauge fields.

\bigskip
{\bf 2.} Let us first briefly show how these ideas work out in the
Abelian case. We start from the well-known Lagrangian of the Maxwell
electromagnetic theory

\begin{equation}
{\cal L}=-\,\frac{1}{4}\,F_{\mu\nu}F^{\mu\nu}\,,
\label{2.1}
\end{equation}

\bigskip\noindent
where the tensor field $F_{\mu\nu}$ is defined as usual

\begin{equation}
F_{\mu\nu}=\partial_\mu A_\nu-\partial_\nu A_\mu\,.
\label{2.2}
\end{equation}

\bigskip\noindent
Let us now suppose that we would like to have massive photons.  If we
directly put a mass term into the Lagrangian like

\begin{equation}
{\cal L}=-\,\frac{1}{4}\,F_{\mu\nu}F^{\mu\nu}
+\frac{1}{2}\,m^2A_\mu A^\mu\,,
\label{2.3}
\end{equation}

\bigskip\noindent
we would have two problems (maybe more): The theory would lose the
gauge invariance and would not be renormalizable any more. Even with
these problems let us rewrite the Lagrangian (\ref{2.3}) with the
help of an auxiliary field as follows

\begin{equation}
{\cal L}=-\,\frac{1}{4}\,F_{\mu\nu}F^{\mu\nu}
-\frac{1}{2}\,j_\mu j^\mu +m\,j_\mu A^\mu\,.
\label{2.4}
\end{equation}

\bigskip\noindent
We observe that the calculation of the equation of motion for $j_\mu$
and its replacement back into (\ref{2.4}) leads to the previous
Lagrangian (\ref{2.3}).

\medskip
The important part of the present development is to look at the
Lagrangian (\ref{2.4}) again but without considering it necessarily
related to (\ref{2.3}). This occurs when we take $j_\mu$ as a
function of another field. In this case, we cannot assume that
(\ref{2.3}) and (\ref{2.4}) are equivalent anymore, even classically.
The interesting point is that the gauge invariance, lost in
(\ref{2.3}), can be restored in the Lagrangian (\ref{2.4}) if we
assume that $j_\mu$ exhibits the following properties: off-shell
divergenceless and gauge invariance. It is necessary to be off-shell
divergenceless in order to compensate the gauge transformation of
$A_\mu$, i.e.

\begin{eqnarray}
\delta A_\mu\,j^\mu&=&\partial_\mu\alpha\,j^\mu\,,
\nonumber\\
&\longrightarrow& -\,\alpha\,\partial_\mu j^\mu\,,
\nonumber\\
&=&0\,,
\label{2.5}
\end{eqnarray}

\bigskip\noindent
where the second step above contains a total derivative.  Concerning
the gauge invariance of $j_\mu$, it is an assumption that can always
be done, in principle for the Abelian case.

\medskip
In order to fulfill these two conditions, tensor fields emerge
naturally by writing $j_\mu$ as a topological quantity
\footnote{We use the convention that the scalar product between two
antisymmetric quantities shall display a factor that takes care of
the multiplicity of terms. This is the reason for the factor 1/2 in
expression (\ref{2.6}). We are also considering that
$\epsilon^{0123}=1=-\,\epsilon_{0123}$ and that the flat metric tensor
reads $\eta_{\mu\nu}=(+,-,-,-)=\eta^{\mu\nu}$.}

\begin{equation}
j_\mu=\frac{1}{2}\,\epsilon_{\mu\nu\rho\lambda}\,
\partial^\nu B^{\rho\lambda}\,.
\label{2.6}
\end{equation}

\bigskip\noindent
We assume that $B^{\mu\nu}$ is independent of the gauge
transformation of the vector field, characterized by the parameter
$\alpha(x)$ above.  Consequently, the gauge invariance condition for
the topological current $j_\mu$ is directly verified. On the other
hand, the antisymmetric tensor field can have its own gauge
transformation.  Using the vector parameter $\xi_\mu(x)$ to
characterize it, we have

\begin{equation}
\delta B^{\rho\lambda}(x)=\partial^\rho\xi^\lambda(x)
-\partial^\lambda\xi^\rho(x)\,.
\label{2.7}
\end{equation}

\bigskip\noindent
The gauge transformations given by (\ref{2.7}) are not all
independent. We notice that $\delta B^{\rho\lambda}=0$ if $\xi^\rho$
is replaced by the derivative of some scalar quantity. We also verify
that $j_\mu$ remains invariant for the gauge transformation given by
(\ref{2.7}).  If we assume that $A_\mu$ does not depend on it, the
Lagrangian (\ref{2.4}) will be invariant for these two gauge
transformations.

\bigskip
Summarizing all the results above, we have

\begin{eqnarray}
(i)&&{\rm Vector\,\,gauge\,\,transformations:}
\nonumber\\
&&\delta_\alpha A_\mu=\partial_\mu\alpha\,,
\nonumber\\
&&\delta_\alpha B_{\mu\nu}=0\,,
\nonumber\\
&&\delta_\alpha j_\mu=0\,,
\label{2.8}\\
\nonumber\\
(ii)&&{\rm Tensor\,\,gauge\,\,transformations:}
\nonumber\\
&&\delta_\xi B_{\mu\nu}=\partial_\mu\xi_\nu
-\partial_\nu\xi_\mu\,,
\nonumber\\
&&\delta_\xi A_\mu=0\,,
\nonumber\\
&&\delta_\xi j_\mu=0\,,
\label{2.9}
\end{eqnarray}

\bigskip\noindent
We have used different subscripts to denote both transformations.
Making now the replacement of $j_\mu$, given by (\ref{2.6}), into the
Lagrangian (\ref{2.4}), we obtain

\begin{eqnarray}
-\,\frac{1}{2}\,j_\mu\,j^\mu
&=&-\,\frac{1}{8}\,\epsilon_{\mu\nu\rho\lambda}\,
\epsilon^{\mu\zeta\alpha\beta}\,\partial^\nu B^{\rho\lambda}
\partial_\zeta B_{\alpha\beta}\,,
\nonumber\\
&\equiv&-\,\frac{1}{72}\,\epsilon_{\mu\nu\rho\lambda}\,
\epsilon^{\mu\zeta\alpha\beta}\,
H^{\nu\rho\lambda}H_{\zeta\alpha\beta}\,,
\nonumber\\
&=&\frac{1}{12}\,H_{\mu\nu\rho}\,H^{\mu\nu\rho}\,,
\label{2.10}
\end{eqnarray}

\bigskip\noindent
where the tensor $H_{\mu\nu\rho}$ is defined as

\begin{equation}
H_{\mu\nu\rho}=\partial_\mu B_{\nu\rho}
+\partial_\rho B_{\mu\nu}
+\partial_\nu B_{\rho\mu}\,.
\label{2.11}
\end{equation}

\bigskip\noindent
We write down the final expression of the Lagrangian as it usually
appears in literature~\cite{Allen,Kalb,Barc1,Kaul}

\begin{equation}
{\cal L}=-\,\frac{1}{4}\,F_{\mu\nu}F^{\mu\nu}
+\frac{1}{12}\,H_{\mu\nu\rho}H^{\mu\nu\rho}
+\frac{1}{2}\,m\,\epsilon_{\mu\nu\rho\lambda}\,
A^\mu\partial^\nu B^{\rho\lambda}\,.
\label{2.12}
\end{equation}

\bigskip\noindent
It is important to emphasize that the Lagrangian above, although
gauge invariant, effectively describes a massive vector gauge
field~\cite{Allen}. This is achieved, for example, by considering the
path integral formalism and integrating out the tensor fields. An
effective Lagrangian for vector fields is then obtained. Their
propagators present a massive pole with the same mass $m$ of the
classical analysis given by the combination between (\ref{2.3}) and
(\ref{2.4}). For details, see reference~\cite{Barc1}.  It might also
be opportune to mention that the mass generation embodied in
(\ref{2.12}) is symmetrical, that is to say, the elimination of
vector field gives also mass to the tensor one.

\bigskip
{\bf 3.} In order to implement these ideas in the SW, it is necessary
to have the non-Abelian formulation of the vector-tensor gauge
theory. Let us mention that this is not a trivial
subject~\cite{Tow,Barc2} and we can directly understand why this
occurs. The non-Abelian version of the tensor gauge transformation
(\ref{2.7}) must be

\begin{equation}
\delta B_{\mu\nu}^a=\bigl(D_\mu\xi_\nu\bigr)^a
-\bigl(D_\nu\xi_\mu\bigr)^a\,.
\label{3.1a}
\end{equation}

\bigskip\noindent
Here we notice that if we replace the gauge parameter $\xi_\mu^a$ by
a derivative (even covariant) of some spacetime scalar we do not get
zero as in the Abelian case. So, the reducibility condition does not
happen in the non-Abelian formulation. This is the origin of the
problem. A non-Abelian gauge theory is incompatible with the Abelian
limit because there is a discontinuity between the two sectors (the
Abelian case has more symmetries than the non-Abelian one). A
possible solution for this problem is to introduce a kind of
Stuckelberg field in order to make compatible the symmetries of the
two sectors~\cite{Barc2}. However, for our particular purposes in the
present paper, that is just to calculate the masses of free vector
fields, we do not need to know details of higher order interaction
involving vector and tensor fields. These masses are obtained as
being poles of the propagators of the vector fields.

\medskip
Let us then write down the gauge field sector of the SW theory 

\begin{equation}
{\cal L}_g=-\,\frac{1}{4}\,F_{\mu\nu}^aF^{a\,\mu\nu}
-\,\frac{1}{4}\,F_{\mu\nu}F^{\mu\nu}\,,
\label{3.1}
\end{equation}

\bigskip\noindent
where

\begin{eqnarray}
F^a_{\mu\nu}&=&\partial_\mu W_\nu^a-\partial_\nu W_\mu^a
+g\,\epsilon^{abc}\,W_\mu^b\,W_\nu^c\,,
\label{3.2}\\
F_{\mu\nu}&=&\partial_\mu B_\nu-\partial_\nu B_\mu\,.
\label{3.3}
\end{eqnarray}

\bigskip\noindent
Here, $W_\mu^a$ are the gauge fields related to the $SU(2)_L$
symmetry and $B_\mu$ to the $U(1)$ hypercharge. This last one is a
combination between electromagnetic and neutral weak fields. The same
occurs with $W_\mu^3$. These combinations are expressed in terms of
the Weinberg angle $\theta_W$ as follows

\begin{eqnarray}
B_\mu&=&\cos\theta_W\,A_\mu-\sin\theta_W\,Z_\mu\,,
\nonumber\\
W_\mu^3&=&\sin\theta_W\,A_\mu+\cos\theta_W\,Z_\mu\,.
\label{3.4}
\end{eqnarray}

\bigskip

In order to obtain mass for gauge fields, we follow the same
procedure of the Abelian case and introduce the Lagrangian

\begin{equation}
{\cal L}_j=-\,\frac{1}{2}j_\mu^a\,j^{a\mu}
-\,\frac{1}{2}j_\mu\,j^{\mu}+M\,j_\mu^a\,W^{a\mu}
+M^\prime\,j_\mu\,B^\mu+\dots\,,
\label{3.5}
\end{equation}

\bigskip\noindent
where dots are representing the remaining terms related to the
non-Abelian formulation~\cite{Tow,Barc2}. Classically, if one
eliminates $j_\mu$ and $j_\mu^a$ by using their equation of motion,
the mass terms $\frac{1}{2}MW_{a\mu}W^{a\mu}$ and
$\frac{1}{2}M^\prime B_\mu B^\mu$ will be obtained. the same occurs
in the quantum point of view, when tensor gauge fields are
introduced~\cite{Barc1}. Since the mass poles obtained in the quantum
propagators are the same of the classical formalism, when tensor
gauge fields are eliminated, we continue to work classically
throughout the paper. This avoid us to run into desnecessary
algebraic complications.

\medskip
Of course, we do not want a mass generation like that, where the
photon field also becomes massive. Let us then use the combination
given by (\ref{3.4}) into the last two terms of (\ref{3.5}). The
result is

\begin{eqnarray}
M\,j_\mu^3\,W^{3\mu}+M^\prime\,j_\mu\,B^\mu
&=&\bigl(M\sin\theta_W\,j^{3\mu}
+M^\prime\cos\theta_W\,j^\mu\bigr)\,A_\mu
\nonumber\\
&+&\bigl(M\cos\theta_W\,j^{3\mu}
-M^\prime\sin\theta_W\,j^\mu\bigr)\,Z_\mu\,.
\label{3.6}
\end{eqnarray}

\bigskip\noindent
Since we do not want a mass generation for the photon field, we have
that $j_\mu$ and $j_\mu^3$ cannot be independent. We thus take

\begin{equation}
M\sin\theta_W\,j^{3\mu}+M^\prime\cos\theta_W\,j^\mu=0\,.
\label{3.7}
\end{equation}

\bigskip\noindent
This permit us to also eliminate the topological current $j^\mu$. Hence,

\begin{eqnarray}
-\,\frac{1}{2}j_\mu^3\,j^{3\mu}
&\!\!\!-\!\!\!&\frac{1}{2}j_\mu\,j^{\mu}+M\,j_\mu^3\,W^{3\mu}
+M^\prime\,j_\mu\,B^\mu
\nonumber\\
&\!\!\!=\!\!\!&-\,\frac{1}{2}\,
\Bigl[1+\Bigl(\frac{M}{M^\prime}\Bigr)^2
\tan^2\theta_W\Bigr]\,j_\mu^3\,j^{3\mu}
+\,\frac{M}{\cos\theta_W}\,j^3_\mu\,Z^\mu
\label{3.8}
\end{eqnarray}

\bigskip\noindent
The equation for $j^3_\mu$ then reads

\begin{equation}
j^3_\mu=\frac{\frac{\strut\displaystyle M}
{\strut\displaystyle\cos\theta_W}}{1+\strut\displaystyle
\Bigl(\frac{\strut\displaystyle M}{\strut\displaystyle M^\prime}
\Bigr)^2\tan^2\theta_W}\,Z_\mu
\label{3.9}
\end{equation}

\bigskip\noindent
Replacing it back into (\ref{3.8}), we get

\begin{eqnarray}
-\,\frac{1}{2}j_\mu^3\,j^{3\mu}
&\!\!\!-\!\!\!&\frac{1}{2}j_\mu\,j^{\mu}+M\,j_\mu^3\,W^{3\mu}
+M^\prime\,j_\mu\,B^\mu
\nonumber\\
&\!\!\!=\!\!\!&-\,\frac{1}{2}\,
\frac{\Bigl(\frac{\strut\displaystyle M}
{\strut\displaystyle\cos\theta_W}\Bigr)^2}
{1+\strut\displaystyle\Bigl(
\frac{\strut\displaystyle M}{\strut\displaystyle M^\prime}
\Bigr)^2\tan^2\theta_W}\,Z_\mu Z^\mu
\label{3.10}
\end{eqnarray}

\bigskip\noindent
We thus observe that the mass generated for $Z_\mu$ reads

\begin{equation}
M_Z=\frac{\strut\displaystyle M}{\strut\displaystyle\cos\theta_W\,
\sqrt{1+\Bigl(\frac{\strut\displaystyle M}
{\strut\displaystyle M^\prime}^2\Bigr)\,\tan^2\theta_W}}
\label{3.11}
\end{equation}

\bigskip\noindent
where $M$ is the mass of $W_\mu^1$ and $W_\mu^2$. Here, $M^\prime$ is
a free parameter that has to be conveniently fixed, accordingly the
experiment. Since we already know that $M$ and $M_Z$ are related by
$M_Z=M/\cos\theta_W$, we conclude that $M^\prime$ must be an infinite
parameter (or $M^\prime\gg M$). Our procedure is then in agreement
with the results of the spontaneous symmetry breaking of the SW
theory. But we emphasize that there are no Higgs bosons here.

\bigskip
{\bf 4.} In conclusion, we have used an alternative mechanism of
mass generation for gauge fields in the SW theory, without Higgs,
that is based on a vector-tensor gauge theory where vector and
tensor fields are coupled in a topological way. The physical
interpretation of our result is that the massive vector fields, that
we effectively see in nature, must be related to massless vector and
tensor gauge fields at some stage of the same nature. We would like
to stress that this might not be an isolated fact, just considered to
be an alternative mass generation mechanism.  Antisymmetric tensor
degrees of freedom might be the reason for the intriguing spacetime
dimension $D=10$ of superstring theories. There is a possibility of
anomaly cancellation in these theories for $D=4$ if antisymmetric
tensor degrees of freedom are introduced~\cite{Barc3}.

\medskip
Unfortunately, there remains an open question. The present mechanism
does not appear to be appropriate to generate mass for matter fields.
We guess that this point requires a better comprehension of the role
played by the masses of the matter fields in the context of the SW
theory. This point is presently under study and possible results
shall be reported elsewhere~\cite{Barc4}

\vspace{1cm}
\noindent {\bf Acknowledgment:} 
This work is supported in part by Conselho Nacional de
Desenvolvimento Cient\'{\i}fico e Tecnol\'ogico - CNPq, Financiadora
de Estudos e Projetos - FINEP and Funda\c{c}\~ao Universit\'aria
Jos\'e Bonif\'acio - FUJB (Brazilian Research Agencies).

\end{document}